\documentclass[twocolumn,superscriptaddress]{revtex4-2}%
\usepackage{amsmath}%
\usepackage{amsfonts}%
\usepackage{amssymb}%
\usepackage{graphicx}
\usepackage{fullpage}
\usepackage{dsfont}
\usepackage{bm}
\usepackage{multirow}
\usepackage{color}
\usepackage{float}
\usepackage{braket}
\usepackage[utf8x]{inputenc}

\begin{document}

%\title{Exciton-polariton optics in atomically thin semiconductors}
\title{Signatures of dark excitons in exciton-polariton optics of transition metal dichalcogenides}
%\title{Signatures of dark excitons in exciton-polariton optics of atomically-thin semiconductors}

\author{Beatriz Ferreira}
\email{beatriz.ferreira@chalmers.se}
\affiliation{Chalmers University of Technology, Department of Physics, 412 96 Gothenburg, Sweden}

\author{Roberto Rosati}
\affiliation{Department of Physics, Philipps-Universit\"at Marburg, Renthof 7, D-35032 Marburg, Germany}

\author{Jamie M. Fitzgerald}
\affiliation{Department of Physics, Philipps-Universit\"at Marburg, Renthof 7, D-35032 Marburg, Germany}

\author{Ermin Malic}
\affiliation{Department of Physics, Philipps-Universit\"at Marburg, Renthof 7, D-35032 Marburg, Germany}
\affiliation{Chalmers University of Technology, Department of Physics, 412 96 Gothenburg, Sweden}

\begin{abstract}
Integrating 2D materials into high-quality optical microcavities opens the door to fascinating many-particle phenomena including the formation of exciton-polaritons. These are hybrid quasi-particles inheriting properties of both the constituent photons and excitons. In this work, we investigate the so-far overlooked impact of dark excitons on the momentum-resolved absorption spectra of hBN-encapsulated WSe$_2$ and MoSe$_2$ monolayers in the strong-coupling regime. In particular, thanks to the efficient phonon-mediated scattering of polaritons into energetically lower dark exciton states, the absorption of the lower polariton branch in WSe$_2$ is much higher than in MoSe$_2$. It shows unique step-like increases in the momentum-resolved profile indicating opening of specific scattering channels. We study how different externally accessible quantities, such as temperature or mirror reflectance, change the optical response of polaritons. Our study contributes to an improved microscopic understanding of exciton-polaritons and their interaction with phonons, potentially suggesting experiments that could determine the energy of dark exciton states via momentum-resolved polariton absorption.
\end{abstract}

\maketitle

\section{Introduction}\label{sec1}

Monolayers of transition metal dichalcogenides (TMDs) show a rich exciton landscape, including bright and dark exciton states \cite{Mueller18,Berghauser18}. This class of atomically-thin materials exhibits a large oscillator strength and exciton binding energies in the range of a few hundreds of meV, hence governing the optoelectronic properties even at room temperature \cite{he2014,ugeda2014,Wang18,Mueller18,brunetti2018}. TMDs have already been successfully integrated into optical cavities \cite{liu2015,dufferwiel2015,schneider2018}, where the coupling of cavity photons with excitons gives rise to the formation of exciton-polaritons \cite{hopfield1958theory,deng2010exciton}. It is only the bright exciton states that can couple to photons to form these quasi-particles, while momentum-dark excitons \cite{Mueller18, Malic18, Berghauser18} cannot be directly accessed by light. Nonetheless, these states are available scattering partners for polaritons via the interaction with phonons. The presence of dark excitons is expected to significantly change the polariton-phonon scattering rates, especially for tungsten-based TMDs, since here dark excitons are the energetically lowest states \cite{Zhang15,Selig16,Brem20, Rosati21d}. So far, polariton-phonon interactions in TMDs have not been well studied, leaving many open questions on the impact of dark exciton states on polariton absorption. \\

\begin{figure}[t!]
\centering
	\includegraphics[width=0.75\columnwidth]{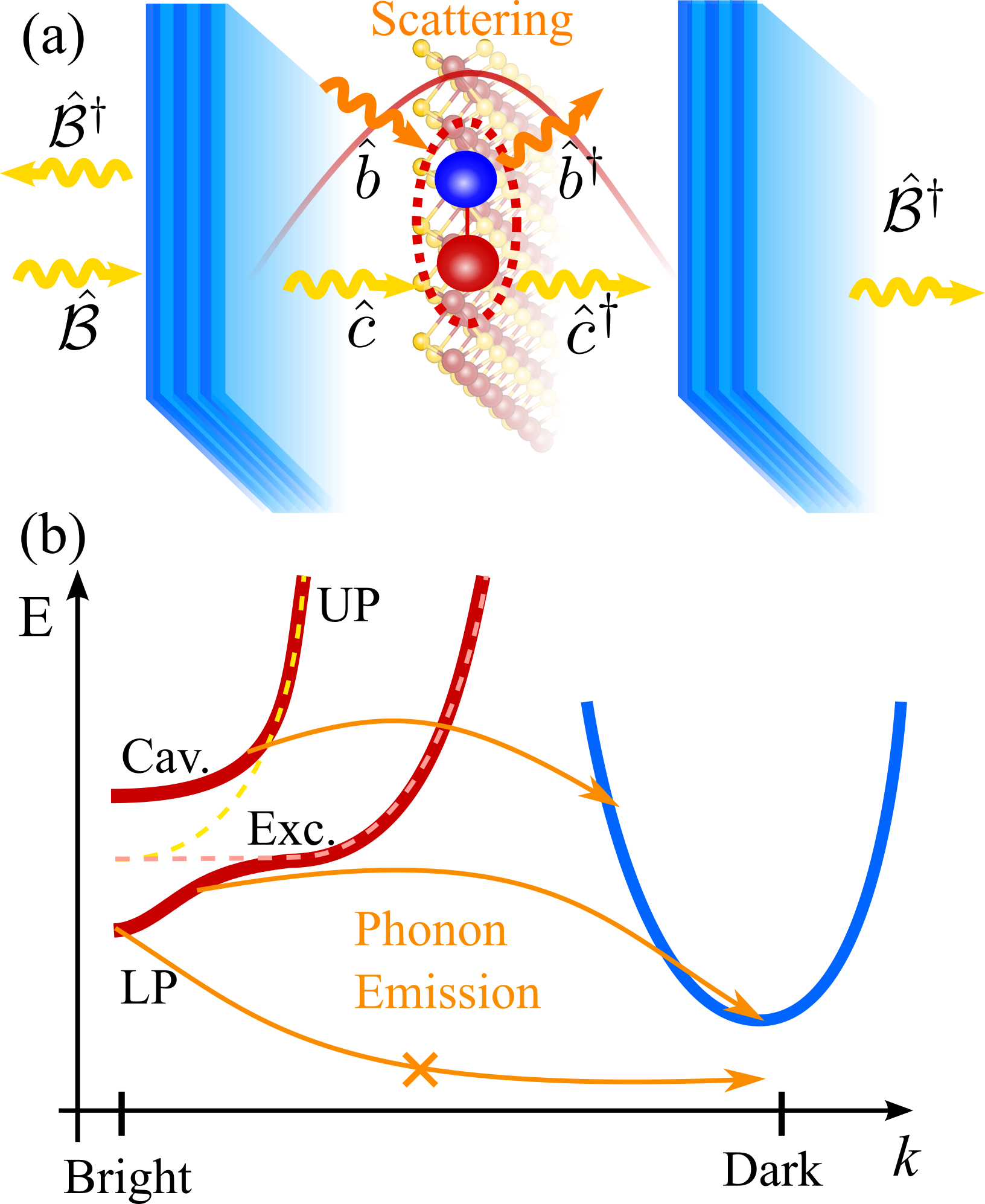}
	\caption{(a) Schematic illustration of a TMD monolayer in a Fabry-Perot cavity with the fundamental cavity mode represented by the red curve. TMD excitons interact with photons and phonons as indicated by the creation (annihilation) operators for photons ($\hat{c}^\dagger(\hat{c})$) and phonons ($\hat{b}^\dagger(\hat{b})$). The cavity system interacts with the outside world via the operators 	$\hat{\mathcal{B}}^\dagger(\hat{\mathcal{B}})$. (b) Exciton-polariton band structure, where polaritons can scatter into dark exciton states by emitting phonons if momentum and energy can be conserved.}
	\label{fig_1}
\end{figure}

In our previous work, we studied  transport properties of exciton-polaritons in MoSe$_2$ monolayers \cite{Ferreira2022}, where dark excitons do not play an important role as they are energetically higher than the 1s bright states \cite{Selig16,Mueller18,Malic18}. Now, the focus lies on the optical response of polaritons in WSe$_2$ monolayers, specifically addressing the impact of dark exciton states. In this regard, the polariton absorption is especially informative as it unambiguously demonstrates strong coupling via the Rabi splitting \cite{kavokin2017}, and its magnitude is determined by the balance between the polariton-phonon and cavity decay rates. We microscopically calculate the polariton absorption by combining the Heisenberg-Langevin equations \cite{gardiner1985} for polaritons with the exciton density matrix formalism \cite{Selig18, Brem20c}. We calculate the full valley- and momentum-dependent polariton-phonon scattering rates that govern the optical response of TMD materials via both spectral linewidths and magnitude. In particular, we explore this in the context of the critical coupling condition \cite{haus1984}, where the total cavity decay rate coincides with the polariton-phonon scattering rate. We predict that the presence of dark excitons has a large impact on the polariton scattering rates, giving rise to clear signatures in momentum-resolved absorption spectra that could be exploited to measure the energy of dark exciton states. Furthermore, we predict and explain a surprising difference in absorption intensity between the upper and lower polariton branch at zero momentum and zero detuning, despite equal photonic and excitonic contributions. We also study the influence of externally accessible quantities to tune the scattering rates (via temperature) and cavity decay rates (via mirror reflectance). For the latter, we find that the cavity quality factor plays an important role for the absorption, in particular for the lower polariton branch that has a smaller photonic component.

\section{Theory}\label{sec2}
We start by describing the theoretical approach to microscopically calculate the absorption spectrum for polaritons in TMD monolayers integrated into an optical cavity. Exciton energies and wavefunctions in TMD monolayers are obtained by solving the Wannier equation \cite{Haug09test, berghauser14, Selig16} including DFT input on single-particle energies \cite{Kormanyos15}. TMDs are characterized by regular bright excitons that are directly accessible in optical spectra, as well as dark exciton states that are known to be the energetically lowest states in tungsten-based TMDs \cite{Kormanyos15,Zhang15,Deilmann2019, Selig16,Brem20, Rosati21d}. In this work, we focus on momentum-dark excitons consisting of Coulomb-bound electrons and holes that are located at different valleys within the Brillouin zone ($K$, $K^\prime$ or $\Lambda$). This means that the required large momentum transfer cannot be provided by photons, making these states optically dark \cite{feierabend2017, Brem20,Malic18,Deilmann2019,erkensten2021, wallauer2021}.\\ 

In this work, we combine the density matrix formalism with the Hopfield approach \cite{hopfield1958theory}, to model the optical response of polaritons. We quantize separately a single internal cavity mode of a Fabry-Perot resonator and the external radiation fields, which are split into two sets of continuum modes corresponding to the left and the right of the cavity (Fig.\ref{fig_1}). The internal and external modes are weakly coupled via the end mirrors, where the in-plane wavevector is conserved. The starting point is the many-particle Hamiltonian in the excitonic picture $\hat{H}=\hat{H}_0+\hat{H}_{X-c}+\hat{H}_{X-b}+\hat{H}_{\mathcal{B}-c}$.
The first term reads in second quantization
\begin{align}
\nonumber
\hat{H}_0&=\sum_{v\mathbf{k}}E^\text{X}_{v,\mathbf{k}}\hat{X}^\dagger_{v\mathbf{k}}\hat{X}_{v\mathbf{k}}+\sum_{\mathbf{k}}E^\text{c}_{ \mathbf{k}}\hat{c}^\dagger_{\mathbf{k}}\hat{c}_{\mathbf{k}}+\sum_{ \mathbf{q}}E^\text{b}_{\alpha \mathbf{q}}\hat{b}^\dagger_{\alpha\mathbf{q}}\hat{b}_{\alpha\mathbf{q}}\\
&+\sum_{j=L,R}\sum_{\mathbf{k}}\int_0^\infty d\omega\,\hbar\omega(\mathbf{k})\hat{\mathcal{B}}^\dagger_{j\mathbf{k}\omega}\hat{\mathcal{B}}_{j\mathbf{k}\omega}
\end{align}
 and describes the free energy of excitons $E^\text{X}_{v\mathbf{k}}$, phonons $E^\text{b}_{\alpha \mathbf{q}}$ as well as photons within ($E^\text{c}_{\mathbf{k}}$) and outside the cavity ($\hbar \omega$).
 Here, $v$ is the exciton  index (we consider only 1s states), $\alpha$ the phonon mode, $\mathbf{k}$ and  $\mathbf{q}$ are the in-plane momentum of excitons/photons (center-of-mass momentum for excitons) and  phonons, respectively. Furthermore, we have introduced $\hat{X}^\dagger_{v\mathbf{k}}(\hat{X}_{v\mathbf{k}})$, $\hat{b}^\dagger_{\alpha\mathbf{q}}(\hat{b}_{\alpha\mathbf{q}})$, $\hat{c}^\dagger_\mathbf{k}(\hat{c}_\mathbf{k})$, $\hat{\mathcal{B}}^\dagger_{j\mathbf{k}\omega}(\hat{\mathcal{B}}_{j\mathbf{k}\omega})$  as exciton, phonon, inner-cavity and outer-cavity photon creation (and annihilation) operators, respectively.\\
 The second term in the Hamiltonian, $\hat{H}_{X-c}=\sum_{v\mathbf{k}} g_{\mathbf{k}}\left(\hat{c}^\dagger_{\mathbf{k}} \hat{X}_{v\mathbf{k}}+\hat{c}_{\mathbf{k}}\hat{X}^\dagger_{v\mathbf{k}}\right)$ describes the exciton-light interaction mediated by the exciton-photon coupling matrix element $g_{\mathbf{k}}$ \cite{Selig16,Brem18}, where photons need to have the same in-plane momentum $\mathbf{k}$ as excitons to fulfill the momentum conservation (hence restricting the coupling only to the bright exciton states). In general, the out-of-plane component $k_z$ influences the cavity energy and exciton-photon coupling. However, we assume the existence of one resonant photon mode (i.e., $E^\text{X}_{\text{KK},0}=E^\text{c}_0$). 
  The third contribution in the Hamiltonina $\hat{H}_{X-b}=\sum_{vv' \mathbf{k} \alpha \mathbf{q}}\mathcal{D}^{v v'}_{\alpha \mathbf{q}}\hat{X}^\dagger_{v\mathbf{k}+\mathbf{q}}\hat{X}_{v'\mathbf{k}}(\hat{b}^\dagger_{\alpha,-\mathbf{q}}+\hat{b}_{\alpha\mathbf{q}})$ describes the exciton-phonon interaction \cite{Selig16}, where the coupling strength is determined by the exciton-phonon matrix element $\mathcal{D}^{v v'}_{\alpha\mathbf{q}}$. Finally, the last term, $\hat{H}_{\mathcal  {B}-c}=i\hbar\sum_{j=L,R}\sum_{\mathbf{k}}\int_0^\infty \frac{d\omega}{2\pi}a_{j,\mathbf{k}}(\omega)[\hat{\mathcal{B}}^\dagger_{j\omega\mathbf{k}}\hat{c}_{\mathbf{k}}-\hat{\mathcal{B}}_{j\omega\mathbf{k}}\hat{c}^\dagger_{\mathbf{k}}]$,  provides the interaction between the inner- and outer-cavity photons \cite{gardiner1985,fitzgerald2022}. The free photons interact with the cavity with a coupling parameter, $a_{j,\mathbf{k}}(\omega)$. Assuming broadband end mirrors, it is appropriate to take the first Markov approximation and approximate this parameter as frequency independent \cite{gardiner1985}. This contribution in the Hamiltonian leads to a consistent description of both the radiative decay rate within the cavity as well as the coupling of polaritons to input and output fields.

Now, we investigate the strong-coupling regime, where the exciton-photon coupling strength $g_{\mathbf{k}}$ is larger than (the difference of) cavity and non-radiative exciton decay rates \cite{deng2010exciton}. The new eigenmodes, known as exciton-polaritons, can be obtained by applying a Hopfield transformation of the excitonic Hamiltonian discussed above, yielding \cite{hopfield1958theory, deng2010exciton}
\begin{align}
\hat{H}=& \sum_{\mathbf{k},n}E^{n}_{\mathbf{k}}\hat{Y}^{n \dagger}_{\mathbf{k}}\hat{Y}^{n}_{\mathbf{k}}+H^0_{b}+H^0_\mathcal{B}+\nonumber\\
&+i\hbar\sum_{\mathbf{k},n, j}\int_0^\infty \frac{d\omega}{2\pi}a_{j\mathbf{k}}(\omega)\times\nonumber\\
&\bigg(h^{n}_{C,\mathbf{k}}\hat{\mathcal{B}}^\dagger_{j\mathbf{k}\omega}\hat{Y}^n_{\mathbf{k}}-h^{n*}_{C,\mathbf{k}} \hat{\mathcal{B}}_{j\mathbf{k}\omega}\hat{Y}^{n\dagger}_{\mathbf{k}} \bigg)+\nonumber\\      
&+\sum_{\mathbf{k}\alpha\mathbf{q} nn'}\tilde{\mathcal{D}}^{n^\prime n}_{\mathbf{k}\alpha\mathbf{q}}\left(\hat{b}^\dagger_{\alpha,-\mathbf{q}}+\hat{b}_{\alpha\mathbf{q}}\right)\hat{Y}^{n' \dagger}_{\mathbf{k}+\textbf{q}}\hat{Y}^{n}_{\mathbf{k}}
\label{1}
\end{align}

\begin{align*}
&\hat{H}= \sum_{\mathbf{k},n}E^{n}_{\mathbf{k}}\hat{Y}^{n \dagger}_{\mathbf{k}}\hat{Y}^{n}_{\mathbf{k}}+\sum_{\mathbf{k}\alpha\mathbf{q} nn'}\tilde{\mathcal{D}}^{n^\prime n}_{\mathbf{k}\alpha\mathbf{q}}\left(\hat{b}^\dagger_{\alpha,-\mathbf{q}}+\hat{b}_{\alpha\mathbf{q}}\right)\hat{Y}^{n' \dagger}_{\mathbf{k}+\textbf{q}}\hat{Y}^{n}_{\mathbf{k}}\nonumber\\
&+i\hbar\sum_{\mathbf{k},n, j}\int_0^\infty \frac{d\omega}{2\pi}a_{j\mathbf{k}}(\omega)\bigg(h^{n}_{C,\mathbf{k}}\hat{\mathcal{B}}^\dagger_{j\mathbf{k}\omega}\hat{Y}^n_{\mathbf{k}}-h^{n*}_{C,\mathbf{k}} \hat{\mathcal{B}}_{j\mathbf{k}\omega}\hat{Y}^{n\dagger}_{\mathbf{k}} \bigg)
\label{1}
\end{align*}
Here, the first term provides the free polaritonic Hamiltonian with $\hat Y^{n\dagger}_{\mathbf{k}} (\hat Y^n_{\mathbf{k}})$ denoting the polariton creation (annihilation) operator  with the polariton mode $n$  and momentum $\mathbf{k}$. 
The energy of the corresponding polariton, $E^n_{\mathbf{k}}$, includes in particular lower and upper polariton branches (LP, UP) that are separated  in $k=0$ by the Rabi splitting $\hbar \Omega_R=E^{UP}_0-E^{LP}_0$.
This is a consequence of the mixing between excitons and photons (with the same center-of-mass and total momentum), as quantified by the Hopfield coefficients \cite{deng2010exciton}. We include also, for notation convenience, polaritons steaming from momentum-dark excitons, although these show no exciton-photon mixing. Nevertheless, we will show below their crucial role for the polariton absorption via additional phonon-induced scattering channels to the optically active polaritons.
Both polariton energies $E^{n}_{\mathbf{k}}$ and Hopfield coefficients $h^n_{X,\mathbf{k}}$ and $h^n_{c,\mathbf{k}}$  are calculated analytically  (with subscript $X$ and $c$ referring to exciton and intra-cavity photon component, respectively) \cite{deng2010exciton}.

The second and the third term in Eq. (\ref{1}) are the free phonon and free outer-cavity photon contribution, respectively, which are not affected by the Hopfield transformation. 
The fourth term describes the interaction of polaritons with the outer-cavity photons, mediated by the photonic Hopfield coefficients as only the photonic part of polaritons couples to the external radiation field. Finally, the last term in Eq. (\ref{1}) describes the polariton-phonon interaction.
Here, the matrix element $\tilde{D}$ is related to the exciton-phonon coupling via $\tilde{\mathcal{D}}^{n^\prime n}_{\mathbf{k}\alpha\mathbf{q}}=h^{n' *}_{\text{X},\mathbf{k}+\mathbf{q}}\mathcal{D}^{n' n}_{\alpha \mathbf{q}}h^{n}_{\text{X},\mathbf{k}}$ and depends on the excitonic Hopfield coefficients $h_X$ \cite{Lengers21}, since phonons only couple to the excitonic part of polaritons. 

To obtain an expression for the polariton absorption, we exploit the Heisenberg equations of motion for the coherent population of polariton and external radiation field (cf. the supplementary information).
For this we make a correlation expansion including the dynamics of the phonon-assisted polarization.
We use the input-output method \cite{gardiner1985} to couple the dynamics between intra- and outer-cavity photon modes at each port. 
We treat the scattering with phonons within a Markov approximation and assuming a thermalized reservoir of incoherent phonons  \cite{Brem20}. 
The absorption then follows from energy conservation as the difference between incoming fields and the total reflected and transmitted light. To simplify the resulting expression, we assume that the cavity is symmetric and ignore interference effects between polaritons in different branches. The latter is a good approximation if the branches are widely spaced in energy compared to the polaritonic spectral width. We obtain an Elliot-like formula for the polariton absorption \cite{fitzgerald2022},
\begin{equation}
A^n_{\mathbf{k}}(\hbar\omega)=\frac{4\gamma_\mathbf{k}^n\Gamma^n_{\mathbf{k}}}{(\hbar\omega-E^n_{\mathbf{k}})^2+(2\gamma^n_\mathbf{k}+\Gamma^n_{\mathbf{k}})^2} \quad ,
\label{abs}
\end{equation}
for each polariton branch and momentum $n,\textbf{k}$. The obtained equation is similar to the expression found in Ref. \cite{fitzgerald2022}, however, the key difference lies in the microscopic treatment of polariton-phonon interaction.  This means that  phonons can change the momentum of the excitonic component of the polariton, leading to a momentum dependent scattering rate. In Eq. (\ref{abs}) we introduced the decay rates
\begin{align}
&\gamma^n_\mathbf{k}=\hbar c(1-|r_m|^2)|h_{\text{c},\mathbf{k}}|^2/(4L_{cav})\label{eqdecayg}\\
&\Gamma^n_{\mathbf{k}}=2\pi \sum_{n^\prime\alpha \mathbf{k}^\prime} |\tilde{\mathcal{D}}_{\alpha , \mathbf{k}^\prime-\mathbf{k}}^{ n^\prime n}|^2 \left(\frac{1}{2}\pm\frac{1}{2}+n^{b}_{\alpha,\mathbf{k}^\prime-\mathbf{k}}\right)\times\nonumber\\
&\times L_{\tilde{\gamma}_0}\left(E^{n^\prime}_{\mathbf{k}^\prime}-E^{n}_{\mathbf{k}}\pm E^b_{\alpha,\mathbf{k}^\prime-\mathbf{k}}\right),
\label{eqdecayG}
\end{align}
where $\gamma^n_\mathbf{k}$ is the effective cavity decay rate of one port and $\Gamma^n_{\mathbf{k}}$ is the polariton-phonon scattering rate. Here we are summing over all possible scattering channels from a polariton  $n,\mathbf{k}$ to all possible receiving polaritons $n',\mathbf{k}'$ via interaction with a phonon with mode $\alpha$ and momentum $\mathbf{q}$, such that the overall momentum is conserved. The quality factor of the cavity reads  $Q_f=E^\text{c}_0L_{cav}/[\hbar c(1-|r_m|^2)|)]$, where $r_m$ is the reflectivity of the cavity. In this work, we use the default value of $r_m=0.99$ if not stated otherwise. 
Importantly, we explicitly consider intervalley scattering by including K$^\prime$ and Q$^\prime$ phonons which allow scattering into polaritons  coinciding with KK$^\prime$ and K$\Lambda$ excitons, respectively.
The polariton-phonon rates are calculated within the Markov-Born approximation \cite{thranhardt2000,Brem18} including effects beyond the 
completed-collision limit \cite{Rossi11} by an energy conservation described via a Lorentzian function with a broadening $\tilde{\gamma}_0=0.1$ meV  \cite{Ferreira2022}. \\
%The polariton absorption depends on the cavity properties as well as the Hopfield coefficients. This is the advantage of the Hopfield method: we can derive an analytical expression for the absorption in terms of the Hopfield coefficients, which is essential to model the physics of the polaritons \cite{kavokin2003thin}. We also need these coefficients to know how the phonon-exciton matrix element is modified, which cannot be done in other methods, like the transfer-matrix method.
Crucially, the polaritonic Elliot formula offers insight into how underlying microscopic decay channels manifest in the absorption of light by polaritons, which would not be possible using the more commonly used classical transfer-matrix method \cite{kavokin2017}. Evaluating Eq. (\ref{abs}) at resonance reveals that absorption is maximized when the two effective polariton decay rates are closest in value. It follows that maximum absorption of $0.5$ is possible at the so-called critical coupling condition \cite{adler1960,haus1984} of $2\gamma^n_\mathbf{k}=\Gamma^n_\mathbf{k}$, i.e. when the leakage out of both ports of the cavity is equal to the exciton dissipation rate within the TMD layer in the cavity.  The maximum possible absorption of 50\% is a well-known constraint for mirror-symmetric two-port systems that support a single resonance \cite{botten1997,piper2014}. We expect the presence of dark excitons to significantly increase the polariton-phonon scattering rates in tungsten-based TMDs (as there they are the energetically lowest states). The opening of intervalley scattering channels is expected to strongly impact the balance between the effective radiative coupling and scattering loss, which  should translate into measurable signatures in polariton absorption spectra. 

\begin{figure}[t!]
\centering
	\includegraphics[width=\columnwidth]{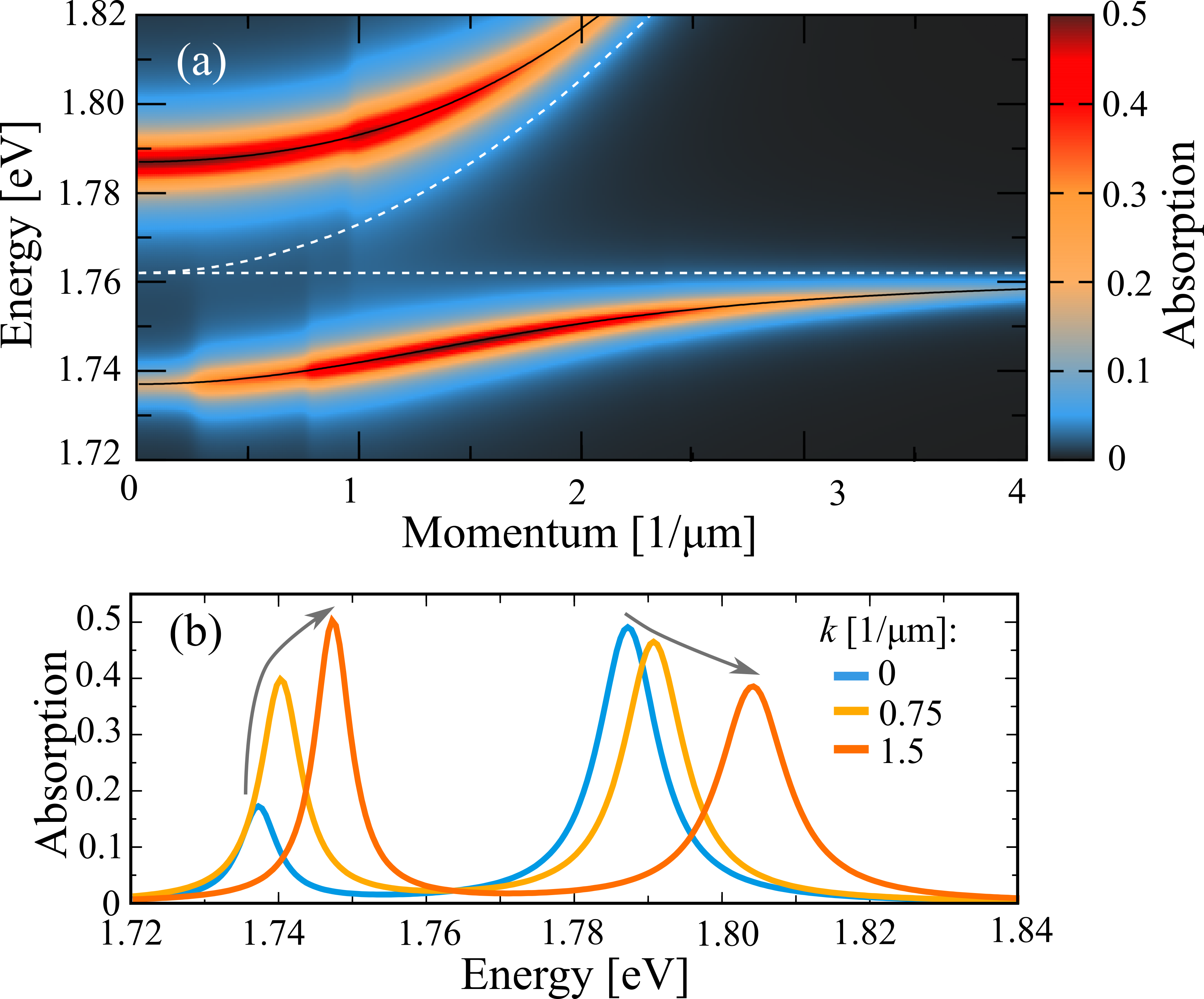}
	\caption{Polariton absorption. (a) Surface plot of absorption in a hBN-encapsulated WSe$_2$ monolayer as a function of momentum and energy at  a temperature of 77 K, assuming a Rabi splitting of $\hbar\Omega_R=50$ meV and a cavity quality factor of $Q_f$=160. The dashed white lines correspond to the bare exciton and cavity dispersion, while the solid black lines describe the polariton dispersion. (b) Absorption cuts as a function of energy for three different momenta.}
	\label{fig_2}
\end{figure}

\section{Results}\label{sec3}

\subsection{Polariton absorption of WSe$_2$ }
Now, we evaluate Eq. (\ref{abs}), using numerically calculated polariton-phonon scattering rates, to study the polariton absorption in the strong-coupling regime for an hBN-encapsulated WSe$_2$ monolayer integrated into a Fabry-Perot cavity with a quality factor of $Q_f\approx160$ and a Rabi splitting of $\hbar\Omega_R=50$ meV. Figure \ref{fig_2}(a) presents an energy- and in-plane momentum-resolved surface plot of the polariton absorption.  
Interestingly, we find the upper polariton to be much higher in intensity than the lower polariton at $k=0$ $\mu$m$^{-1}$ (cf. also the blue lines in Fig.\ref{fig_2}(b)). Previous reports in GaAs have shown that in the case of zero detuning, the lower and upper polariton peaks intensities are similar \cite{chen1995,pau1995}.
In the resonant case, the polaritons have an equal photonic and excitonic contribution at $k=0$, hence also the cavity decay rate is the same for both polaritons. As a result, the phonon-induced decay rate of polaritons must be responsible for the observed difference in the height of absorption peaks. 
Furthermore, we find that the absorption is enhanced for increasing momenta for the lower polariton ($A^{LP}$) up to approximately $k=1.6$ $\mu\text{m}^{-1}$, while it is reduced for the upper polariton ($A^{UP}$), (cf. also the absorption cuts in Fig. \ref{fig_2}(b)). Moreover, we observe that not only the absorption intensity but also the linewidth of $A^{LP}$ becomes larger for increasing in-plane momentum, before it is again reduced for momenta higher than  $k=1.6$ $\mu\text{m}^{-1}$. The absorption intensity and the spectral linewidth of polariton resonances can be ascribed to the interplay of the cavity decay and non-radiative decay of polaritons via scattering with phonons as discussed in detail below.

\subsection{Critical coupling}

\begin{figure}[t!]
\centering
	\includegraphics[width=\columnwidth]{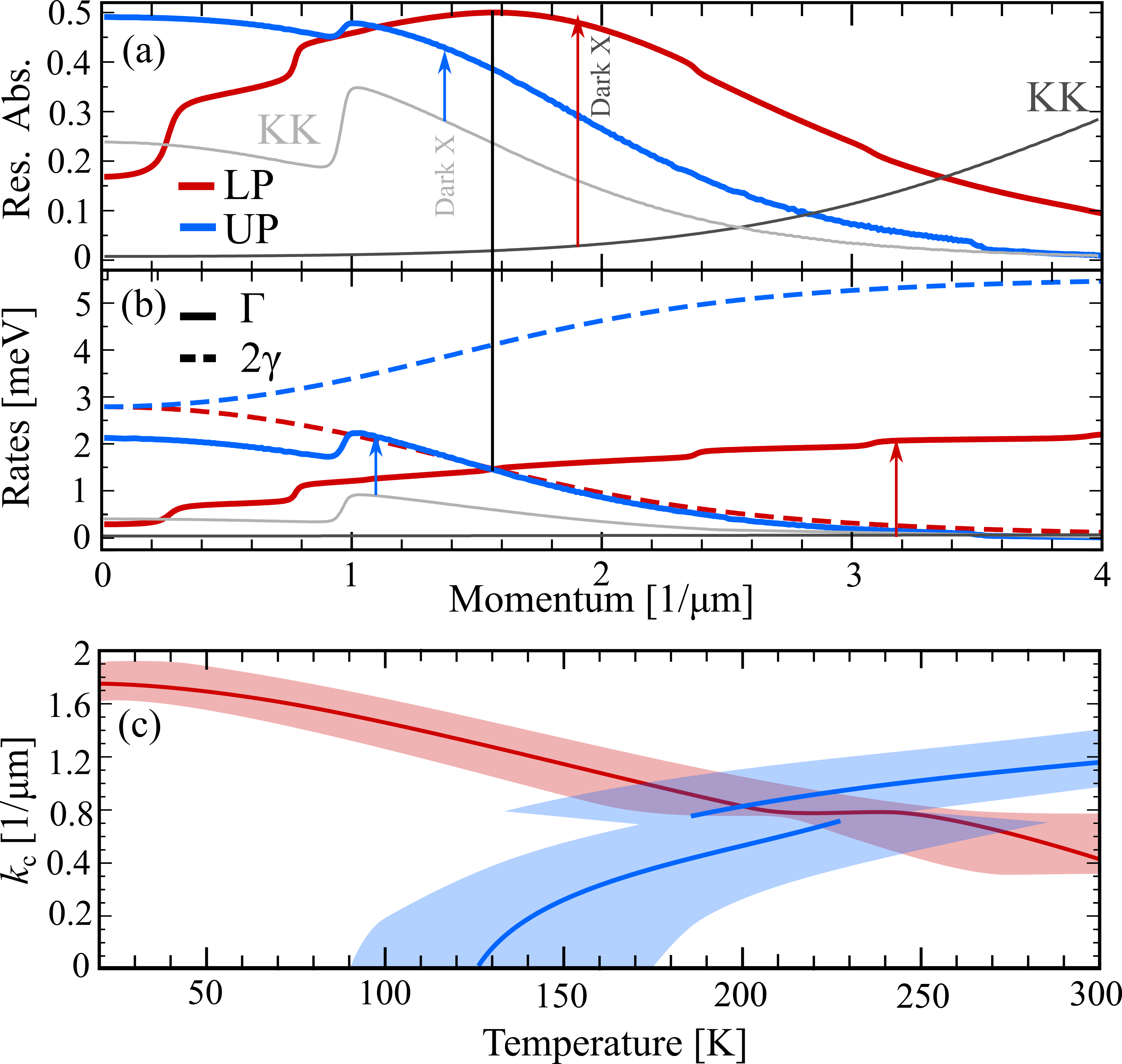}
	\caption{Critical coupling. (a) Maximal absorption at the resonant energy as a function of momentum for the lower (red, LP) and upper (blue, UP) polariton at 77 K and $\hbar\Omega_R=50$ meV, $Q_f\approx$160. (b) Polariton-phonon scattering rate $\Gamma^n_\mathbf{k}$ (solid lines) and cavity decay rate $\gamma^n_\mathbf{k}$ (dashed lines) as a function of momentum for the upper and lower polariton (same colors as in (a)). The maximum value of absorption of $A n=0.5$ identifies the critical coupling conditions $\Gamma^n_\mathbf{k}=2\gamma^n_\mathbf{k}$ for the respective polariton and it is marked by a vertical black line. The grey lines show the case without considering dark states and only taking into account the bright KK excitons. (c) Critical coupling momentum $k_{c}$ as a function of temperature for the upper (blue) and lower polariton (red). The shaded area corresponds to the range $0.5\geq A^n \geq 0.495$. } 
	\label{fig_3}
\end{figure}

To explain the different behavior in the absorption spectra of the upper and lower polariton branch, we plot the maximal absorption $A^n_{\mathbf{k}}$ of the UP and the LP branch at 77 K in Fig. \ref{fig_3}(a). The absorption intensity of the UP branch generally decreases with the momentum, however, with one exception at approximately $k=1$ $\mu$m$^{-1}$, where we observe a small increase (blue line). For the lower polariton branch, then we find, in contrast, an enhanced absorption until approximately $k=1.6$ $\mu$m$^{-1}$ where the critical coupling condition with a maximum possible value of  $A=0.5$, is reached (red line). The increase of the absorption includes several steep step-like enhancements before the absorption starts to decrease for values larger than $k=1.6$ $\mu$m$^{-1}$. 

To better understand the change of the absorption as a function of the in-plane momentum and the opposite behavior of the upper and the lower polariton branch observed in Fig. \ref{fig_3}(a), we investigate the momentum-dependent cavity decay rate $\gamma^n_\mathbf{k}$ and polariton-phonon scattering rate $\Gamma^n_\mathbf{k}$, cf. Eqs. (\ref{eqdecayg}) and (\ref{eqdecayG}). We find that for the lower polariton branch, the critical coupling condition of $\Gamma^n_\mathbf{k}=2\gamma^n_\mathbf{k}$ is reached at $k=1.6$ $\mu$m$^{-1}$, as denoted with the black vertical line in Fig. \ref{fig_3}(b). This corresponds exactly to the momentum where the maximal absorption of $A^{LP}=0.5$ is reached.
The microscopic calculation of polariton-photon scattering rates explains the step-like increase in the absorption of both the UP and LP polariton branch. These can be clearly attributed to an increase of the polariton-phonon scattering rates at certain momenta (at $k\approx 0.3$, 0.8, 2.4 and 3.1 $\mu$m$^{-1}$ for $\Gamma_\mathbf{k}^{LP}$ and at 1 $\mu$m$^{-1}$ for $\Gamma_\mathbf{k}^{UP}$). Importantly, each of these steep increases is a signature of an opening of an intervalley scattering channel into dark exciton states. At momentum $k=0$, the energy $E^{LP}_0$ of the lower polariton is too low to allow scattering into the K$\Lambda$ exciton via emission of phonons (Fig. \ref{fig_1}(b)) as  $E^{LP}_{0}-E^X_{\Lambda,0}\approx 11.2$ meV, which is just smaller than the energy of 11.4 meV of intervalley TA phonons \cite{Jin14}. When $k$ reaches the threshold value of $k\approx 0.3$ $\mu$m$^{-1}$, the scattering channel into $K\Lambda$ states opens, resulting in the abrupt increase of $\Gamma^n_\mathbf{k}$, cf. also the schematic in Fig. \ref{fig_1}(b). Note that, in contrast, the cavity decay rate $\gamma^n_\mathbf{k}$ increases/decreases smoothly with $k$ for the UP/LP branch, cf. the dashed lines in Fig. \ref{fig_3}(b). This increase/decrease is determined by the photonic Hopfield coefficient, which increases for the UP and decreases for the LP branch.

To illustrate the importance of dark excitons, we also show the polariton absorption and the polariton-phonon scattering rates without including dark exciton states, i.e. we only take into account the bright KK excitons (grey lines in Figs. \ref{fig_3}(a,b)). We find that for the lower polariton the resonant absorption is drastically reduced at small momenta, with the critical coupling condition shifted to higher momenta. We also find that the steep increases step-like increases found for these polaritons disappear (red vs. lower grey line), as  they stem from scattering into dark excitons.
For the scattering rates of the lower polariton, the intravalley scattering is orders of magnitude smaller than the intervalley one (grey line is basically 0), due to the forbidden optical absorption for low temperatures and since the scattering of LP polaritons with intravalley acoustic modes is energetically forbidden \cite{Ferreira2022}. In the case of the upper polariton, the qualitative shape of the absorption curve in Fig. \ref{fig_3}(a) is similar to intravalley scattering without dark excitons (blue vs upper grey line). In particular, both lines show a step-like increase at $k\approx 1\mu \text{m}^{-1}$, which stems from the intravalley emission via emission of optical modes.
However, the intensity of the UP absorption is strongly reduced in the absence of dark excitons. This is due to the overall decrease of the polariton-phonon scattering rates, Fig. \ref{fig_3}(b), moving the system further away from the critical coupling condition. 

So far, we have only considered the polariton absorption at 77K, where the critical coupling condition can only be reached for the lower polariton branch. To further investigate this, we present in Fig. \ref{fig_3}(c) the critical coupling momentum $k_{c}$ as a function of temperature for the upper (blue line) and the lower polariton (red line). The blue- and red-shaded areas correspond to the region $0.5\geq A^n \geq 0.495$ to take into account uncertainties in the experimental measurement of the maximal absorption. As we increase the temperature, the critical coupling occurs at smaller momenta for the LP branch due to an overall increase of the scattering with phonons. 
Since the cavity decay rates $\gamma^i_\mathbf{k}$ are temperature-independent within our model, the overall increase in $\Gamma^n_\mathbf{k}$ at higher temperatures results in smaller $k_c$ fulfilling the critical coupling conditions.
Interestingly, for the UP branch, we find that there is no critical coupling for temperatures below approximately 125K.  
We show in Fig. \ref{fig_3}(b) that at $k=0$ the cavity decay rate $\gamma^n_\mathbf{k}$ is larger than $\Gamma^n_\mathbf{k}$. However, while $\gamma^{LP}_\mathbf{k}$ decreases for increasing momenta, thus approaching the smaller values of  $\Gamma^n_\mathbf{k}$, the opposite takes place for $\gamma_\mathbf{k}^{UP}$. Thus, for upper polaritons, the critical coupling can only occur at higher temperatures, where $\Gamma^n_\mathbf{k}$ is considerably enhanced. Interestingly, we find that at around 200K two different momenta fulfill the critical coupling condition for UP (blue lines in Fig. \ref{fig_3}(c)). At these temperatures the cavity decay rate crosses the polariton-phonon scattering rate in the region of the opening of the optical emission (step-like increase), where we can have the same value of scattering and cavity-decay rates for two (or more) momenta.
 
 \subsection{Absorption engineering}
\begin{figure}[t!]
\centering
	\includegraphics[width=\columnwidth]{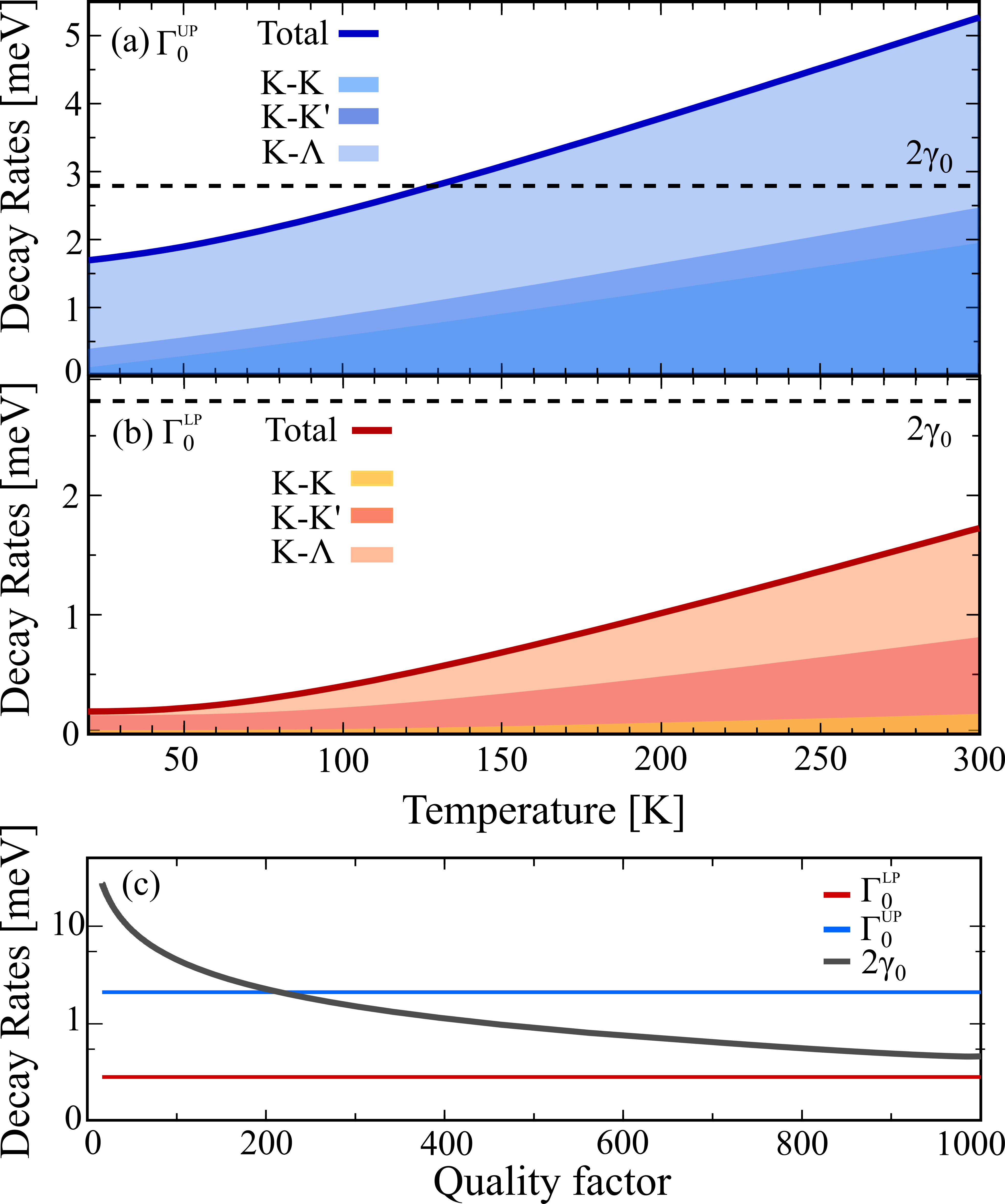}
	\caption{Temperature and quality factor study. Polariton-phonon scattering rate $\Gamma^n_\mathbf{k}$ for the (a) upper (UP) and (b) lower polariton branch (LP) at $k=0$ as a function of temperature (for $Q_f\approx$160). We identify the contributions of  the intravalley (KK) as well as intervalley (KK$^\prime$, K$\Lambda$) scattering channels (shaded areas).  Note that $2\Gamma^i_0$ corresponds to the spectral linewidth of the respective polariton. The crossing point with the   cavity decay rate 2$\gamma^i_0$ (dashed line) corresponds to the critical coupling condition. (c) Polariton-phonon decay rates and the cavity decay rate $\gamma^n_\mathbf{k}$  at $k=0$ as a function of the quality factor (at $T=77$ K).}
	\label{fig_4}
\end{figure}

We have demonstrated that the polariton absorption depends on two key quantities: the polariton-phonon scattering rate $\Gamma^n_\mathbf{k}$ and the cavity decay rate $\gamma^n_\mathbf{k}$. Now, we would like to tune the absorption by changing these quantities. While $\Gamma^n_\mathbf{k}$ is strongly sensitive to temperature, $\gamma^n_\mathbf{k}$ is determined by the cavity quality factor (i.e. in particular the reflectance of the cavity end mirrors).

In Figs. \ref{fig_4}(a) and (b) we show the temperature-dependent polariton-phonon scattering rates at $\mathbf{k}=0$ for the UP and LP branch, respectively. Note that this corresponds to the half linewidth of the respective polariton resonance in absorption spectra. We add up different scattering channels including intravalley scattering within the K valley ($KK$) as well as intervalley scattering into momentum-dark exciton states ($KK^\prime$ and $K\Lambda$). For comparison, the black dashed line shows the temperature-independent $2\gamma_0$, indicating at which temperatures the critical coupling condition is reached at $k_C=0$. This is the case for the UP branch at $T\approx 125$ K - in agreement with Fig. \ref{fig_3}(c). 
In contrast, for the LP branch, the critical coupling at $k_C=0$ can only be reached at temperatures significantly higher than room temperature. 

For both polariton branches, the largest contribution to the linewidth comes from the intervalley scattering into dark $K\Lambda$ excitons reflecting the efficient scattering plus the three-fold degeneracy of the $\Lambda$ valley, similar to the excitonic case \cite{Brem19}. Furthermore, at room temperature, intervalley scattering within the K valley is also important. 
At 20K,  the LP linewidth is determined to a large extent by scattering into the dark $KK^\prime$ excitons. We stress that here we are focusing on the scattering from the $k=0$ polariton state. There are further possible scattering channels at larger momenta, as shown in Fig.\ref{fig_3}(b). 
Increasing the temperature to 300K increases the LP linewidth by around one order of magnitude as the absorption of intervalley phonons becomes possible. 
At 77 K the intravalley contribution to the phonon-scattering rates is very small, in accordance with  Fig.\ref{fig_3}.
The linewidth of the upper polariton is at 20K much larger compared to the LP branch since emission into dark excitons is possible even at $k=0$ thanks to the much higher polariton energy (Fig. \ref{fig_1}(b)). Hence, the increase in UP from 20K to 300K is not as substantial as in the LP case. Overall, Figs. \ref{fig_4}(a,b) illustrate the huge impact of dark exciton states on the polariton-phonon scattering rates and thus on the polariton absorption. 

After having addressed the role of temperature in tuning the polariton-phonon scattering rates $\Gamma^n_\mathbf{k}$, we now focus on the change of the cavity decay rate $\gamma^n_\mathbf{k}$ as a function of the quality factor $Q_f$. In Fig. \ref{fig_4}(c) we show $\gamma^n_\mathbf{k}$ (gray line) and $\Gamma^n_\mathbf{k}$ (red and blue lines) as a function of $Q_f$. We find that UP has a critical coupling condition $k_C=0$ around $Q_f\approx200$. For small values of the quality factor, the UP absorption is expected to increase, but as we move further away from the critical coupling the absorption decreases. For the LP,  the critical coupling condition is only reached at high values of $Q_f$.  

\begin{figure}[t!]
\centering
	\includegraphics[width=\columnwidth]{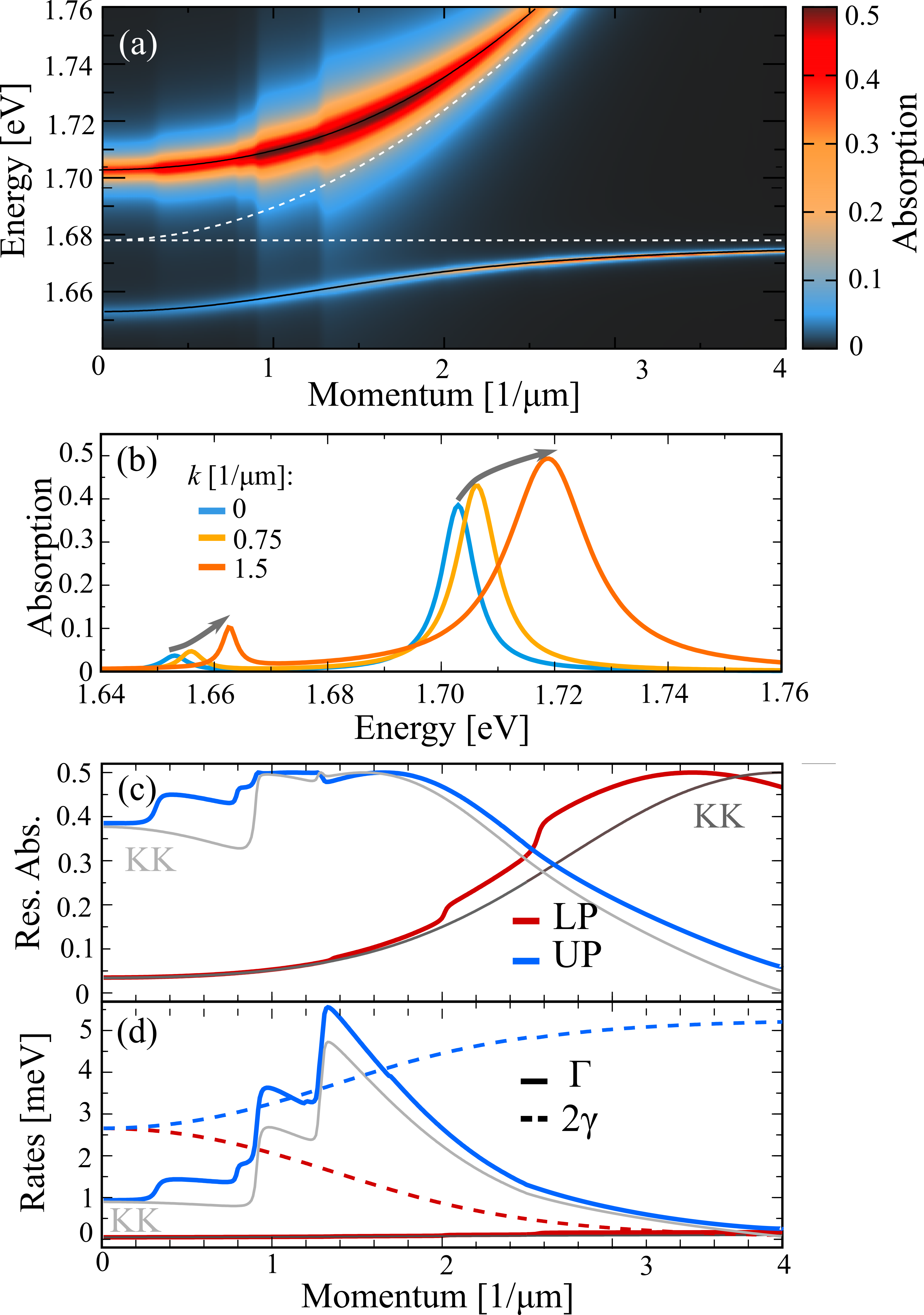}
	\caption{Absorption of MoSe$_2$. (a) Surface plot of polariton absorption of a hBN-encapsulated MoSe$_2$ monolayer as a function of momentum and energy (77K, $\hbar\Omega_R=50$ meV and $Q_f\approx$160).  (b) Absorption cuts as a function of energy for three different momenta. (c) Absorption intensity as a function of  momentum for the lower (LP) and the upper (UP) polariton branch.  (d) Decay rates $\Gamma^n_\mathbf{k}$  and $2\gamma^n_\mathbf{k}$  as a function of momentum for the LP and UP branch.  The thin grey lines in (c) and (d) correspond to the case without dark excitons (considering only the bright $KK$ excitons). }
	\label{fig_5}
\end{figure}

\subsection{Polariton absorption of MoSe$_2$}

So far we have studied the polariton absorption for WSe$_2$ monolayers, where dark excitons turned out to play a crucial role. Now we investigate the MoSe$_2$ monolayer exhibiting a different energetic alignment of dark and bright states. With the latter being the lowest ones in MoSe$_2$ \cite{Mueller18, Selig16,Deilmann2019}, we expect only a negligible contribution from dark excitons.
Similarly to the case of WSe$_2$, we show in Fig. \ref{fig_5}(a) the absorption of polaritons as a function of momentum and energy for the zero-detuning case at $T=$77 K. 
We find a drastic reduction in absorption as well as in the linewidth of the LP absorption compared to WSe$_2$ (Fig. \ref{fig_2}(a)). This can be clearly observed in the momentum cuts shown in Fig. \ref{fig_5}(b). Although the intensity of the resonant absorption increases for larger momenta, similar to the case of WSe$_2$, quantitatively the increase is much slower, reaching only a maximal value of approximately 0.1 at $k=1.5\mu\text{m}^{-1}$ (compared to almost 0.5 predicted for WSe$_2$). 
Interestingly, for larger momenta we find also an increase of the absorption for the UP branch (Fig. \ref{fig_5}(b)) - opposite to the case of  WSe$_2$ (Fig. \ref{fig_2}(b)). In addition, we observe a large increase in the spectral width of polariton resonances at larger in-plane momenta $k$.

To microscopically understand the qualitative as well as quantitative differences of the momentum-resolved absorption in MoSe$_2$ and WSe$_2$, we investigate the intensity of the resonant polariton absorption and the underlying polariton-phonon and cavity decay rates. 
We assume the same value of reflectivity $r_m=0.99$ as in Fig. \ref{fig_2}, resulting in similar cavity decay rates $\gamma^n_\mathbf{k}$ as for WSe$_2$
As a consequence, the observed difference in polariton absorption  must be due to the phonon-scattering rates.
We show both the absorption and decay rates also for the case without dark excitons (grey lines in Figs. \ref{fig_5}(c,d)). As expected, in MoSe$_2$ there is only a minor contribution of dark states. For the LP branch, the scattering into the energetically higher dark exciton states is generally weak as it is driven by absorption of intervalley phonons from high-momenta states. 
Since for LP also the intravalley scattering with acoustic modes is forbidden due to momentum and energy conservation (due to the  almost flat phonon dispersion making it difficult to fulfil the energy conservation \cite{Ferreira2022}), the lower polariton has only a very small scattering rate (the red line in Fig. \ref{fig_4}(d) is almost not visible). 
Nevertheless, the decrease of the cavity decay rate $\gamma^n_\mathbf{k}$ with increasing momenta allows for the critical coupling condition at the very high momenta of $k_c=3.25$ $\mu$m$^{-1}$ (cf. Fig. \ref{fig_5}(d)), where the LP absorption reaches its maximum value of $A^n=0.5$ (Fig. \ref{fig_5}(c)).
Interestingly, even though dark excitons have only a small contribution, their presence shifts the critical coupling condition to a smaller momentum (cf. grey vs red line in Fig. \ref{fig_5}(c)).

For the upper polariton,  the intra-valley scattering contribution is also dominant (only small deviations between the grey and blue line), showing two step-like increases at $k\approx 1$ and $k\approx 1.3 $ $\mu$m$^{-1}$ due to the emission of intravalley optical modes (LO/TO and A1 with energy 36.1/36.6 meV and 30.3 meV, respectively). In contrast to WSe$_2$, we observe a large increase in the phonon-scattering rates for the UP branch, reflecting a more efficient intravalley scattering with optical modes in MoSe$_2$ \cite{Jin14}. This leads to the much broader spectral width of the resonances observed in Fig. \ref{fig_5}(b).
 The contribution of dark excitons is minor, however, we still observe an opening of an emission channel into dark states, cf. the step-like increase of $\Gamma_\mathbf{k}^{UP}$ at $k\approx 0.4$ $\mu$m$^{-1}$.
This opening is important for understanding the increase of the resonant absorption, when going from $k=0$ to $k=0.75$ $\mu$m$^{-1}$ observed in Fig. \ref{fig_5}(b) (in contrast to the prediction for WSe$_2$ in Fig. \ref{fig_1}(a)). Without dark states, there would be a decrease of the absorption up to approximately 0.9  $\mu$m$^{-1}$ (cf. the grey line in Fig. \ref{fig_5}(c)).
 In MoSe$_2$, the upper polariton fulfills the critical coupling condition at the four different momenta $k_{c}\approx$ 1, 1.2, 1.3 and 1.6 $1/\mu$m.  The lowest two are a consequence of polariton scattering into dark exciton states.
%%%%%%%%%%%%%%%%%%%%%%%%%%%%%%%%%%%%%%%%%%%%%%%%%%%%%%%%%%%%%%%%%%%%%%%%
%%%%%%%%%%%%%%%%%%%%%%%%%%%%%%%%%%%%%%%%%%%%%%%%%%%%%%%%%%%%%%%%%%%%%%%%

\section{Conclusion}\label{sec4}
We have studied polariton absorption for hBN-encapsulated WSe$_2$ and MoSe$_2$ monolayers integrated into a Fabry-Perot cavity.
These two materials are ideal to study the impact of momentum-dark exciton states, as they possess an opposite energetic ordering of bright and dark states. Based on a microscopic theory combining the Hopfield approach with an excitonic density matrix formalism, we predict a significant impact of dark excitons on the polariton absorption in WSe$_2$.
This is particularly true for the absorption of the lower-polariton branch, which shows an overall enhancement of the absorption intensity as well as a distinctive step-like increase in momentum-resolved resonant absorption. The latter indicate the opening of scattering channels into dark exciton states, hence potentially suggesting a possibility to measure the energy of dark excitons. Furthermore, we have investigated the critical coupling condition at which a maximum possible absorption intensity can be reached. To tune this condition, we have varied temperature and the cavity quality factor allowing us to control the polariton-phonon scattering rates and cavity decay rates, respectively. Our study provides new microscopic insights into the polariton absorption and the role of dark exciton states and could trigger further experimental studies on exciton-polaritons in atomically-thin materials.

\section*{Acknowledgments}
We thank Marten Richter (TU Berlin) for inspiring discussions. This project has received funding support from the DFG via SFB 1083 (project B9), the European Union's Horizon 2020 Research and Innovation programme under grant agreement no. 881603 (Graphene Flagship) and from the Knut and Alice Wallenberg Foundation via the Grant KAW 2019.0140. The computations were enabled by resources provided by the Swedish National Infrastructure for Computing (SNIC). \\

\bibliography{Main}
\end{document}